\input{aipcheck}

\documentclass[
    ,final            % use final for the camera ready runs
  ]
  {aipproc}

\layoutstyle{6x9}

\begin{document}

\title{Optical observations of GRB 060218/SN 2006aj and its host galaxy}

%\classification{<Replace this text with PACS numbers; choose from this list:
%                \texttt{http://www.aip..org/pacs/index.html}>}

\author{Patrizia Ferrero}{
  address={Th\"uringer Landessternwarte Tautenburg, D-07778 Tautenburg, Germany}
}

\author{Eliana Palazzi}{
  address={INAF -- Istituto di Astrofisica Spaziale e Fisica Cosmica di Bologna, I-40129 Bologna, Italy}
}

\author{Elena Pian}{
  address={INAF -- Osservatorio Astronomico di Trieste, I-34131 Trieste, Italy}
}

\author{Sandra Savaglio}{
  address={Max-Planck-Institut f\"ur Extraterrestrische Physik, D-85748 Garching, Germany \\(on 
  behalf of a larger collaboration)}
}

%\title{on behalf of a larger collaboration}
%\author{on behalf of a larger collaboration}

\begin{abstract}
The supernova SN~2006aj associated with GRB~060218 is the second-closest GRB-SN observed 
to date ($z$=0.033) and is the clearest example of a SN associated 
with a Swift GRB with the earliest optical spectroscopy. Its optical data 
showed that this is the fastest evolving and among the least luminous 
GRB-SNe (70\% as luminous as SN1998bw). However, its expansion velocity and 
a comparison with other stripped-envelope SNe suggest that SN2006aj is an 
intermediate object between Type Ic GRB-SNe and those not accompained by 
a GRB. High-resolution optical spectroscopy together with SDSS pre-burst 
observations revealed that the host galaxy of SN2006aj is a low-luminosity, 
metal-poor star-forming dwarf galaxy.

\end{abstract}

\maketitle

%%%%%%%%%%%%%%%%%%%%%%%%%%%%%%%%%%%%%%%%%%%%
%% MAINMATTER
%%%%%%%%%%%%%%%%%%%%%%%%%%%%%%%%%%%%%%%%%%%%

\section{Introduction}
The first hint for an association of long-duration GRBs with core-collapse supernovae of 
Type Ic \cite{Filippenko1997} came with the contemporaneous discovery of GRB~980425 and 
a local SN (1998bw) in its error circle \cite{Galama1998}. Subsequently, the first 
spectroscopic identification of a SN Ic superposed on a GRB afterglow (GRB~030329/SN~2003dh;
e.g. \cite{Stanek2003}) revealed the unambiguous signatures of this connection. This event was 
followed by GRB~031203/SN~2003lw (e.g. \cite{Malesani2004}). Almost three years later 
a new spectacular case of a nearby GRB-SN association was found: GRB~060218/SN~2006aj.

On 2006 February 18, at 03:34:30 UT, the Burst Alert Telescope (BAT) on board the \emph{Swift}
satellite detected the bright GRB~060218 \cite{Cusumano2006}. The \emph{Swift} X-ray Telescope (XRT) 
and the UV/Optical Telescope (UVOT) also detected its afterglow in the X-ray and optical bands \cite{Cusumano2006,Kennea2006}, respectively, leading to a precise localization of the optical 
counterpart \cite{Marshall2006}. Due to its unusual properties in the gamma-ray band \cite{Campana2006} and
its odd behavior \cite{Gehrels2006}, its nature was not clear till the determination of 
its redshift ($z$= 0.033 \cite{Mirabal2006a}) and the discovery of an association with a supernova \cite{Masetti2006}.
GRB~060218 is classified as an X-ray Flash \cite{Campana2006}, with $E_{iso}$ comparable with 
other GRB-SNe  and consistent with the Amati relation \cite{Amati2006}. It is probably not an off-axis 
event \cite{Nousek2006}.

The low redshift and the brightness of the object allowed extensive follow-up observations with ground-based facilities
\cite{Modjaz2006,Mirabal2006b,Sollerman2006,Pian2006,Ferrero2006,Soderberg2006}.

Here, we summarize results from a photometric and spectroscopic ESO Very Large Telescope (VLT)
campaign on GRB~060218/SN~2006aj, with additional data gathered with the robotic Liverpool 
Telescope (LT), the Lick Observatory 
Shane 3m Telescope and the robotic Katzman Automatic Imaging Telescope (KAIT; 
\cite{Filippenko2001,Li2003}). (For a more detailed description of the observations and of the data reduction see \cite{Pian2006,Ferrero2006}, and references therein.)
In addition we discuss the properties of the host galaxy.

%%%%%%%%%%%%%%%%%%%%%%%%%%%%%%%%%%%%%%%%%%%%

\section{Observations}

The data set includes observations performed over a period of almost 4 weeks by means of different 
telescopes \cite{Pian2006,Ferrero2006}:

\begin{itemize}
\item
  \emph{VLT}-- We observed GRB~060218/SN~2006aj both spectroscopically \cite{Pian2006} and 
  photometrically \cite{Ferrero2006} with the ESO 8.2 m VLT equipped with the FORS1 and the FORS2 
  instruments nearly daily
  from February 21 until 26 days after the burst. Near the epoch of the supernova maximum, we obtained a  
  high-dispersion spectrum with VLT/UVES.
\item
  \emph{Liverpool Telescope}-- Several data were acquired over a period of about 2.5 weeks post-burst 
  by the 2 m robotic Liverpool Telescope on La Palma.
\item
  \emph{Shane and KAIT Telescopes at Lick Observatory}-- An additional spectrum was obtained with 
  the Shane 3 m telescope
  4 days after the burst and other photometric observations were coordinated at the 0.8 m robotic KAIT
  telescope during four consecutive nights between 3 to 6 days after the burst.
\end{itemize}

As the host galaxy and SN~2006aj were not separated at the angular resolution of the images, the host magnitudes (for details see \cite{Ferrero2006} and references therein) have been subtracted from all data. 
For the extinction correction, we used the values derived by \cite{Guenther2006} for our Galaxy and the 
host galaxy, $A_V=0.39$ mag and $A_V=0.13$ mag, respectively.

%%%%%%%%%%%%%%%%%%%%%%%%%%%%%%%%%%%%%%%%%%%%

\section{Results and Discussion}

\subsection{The spectra}

\begin{figure}
  \includegraphics[height=.45\textheight]{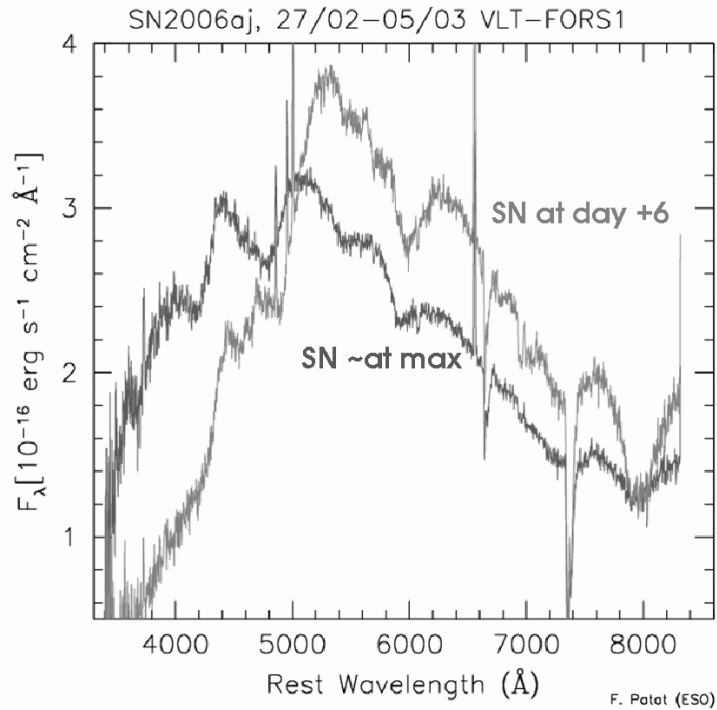}
  \caption{VLT low-resolution spectra of SN~2006aj acquired at its maximum (February 27)
  and six days later (March 5). The strong evolution toward the red is common of Type Ic broad-lined
  SNe and suggests a strong expansion velocity.}
\label{figure1}
\end{figure}

The spectra of SN~2006aj \cite{Pian2006,Mazzali2006}, with their broad absorption lines and their lack of hydrogen and helium
features (Fig. \ref{figure1}), clearly resemble those of broad-lined  Type Ic supernovae. In particular, the deduced properties of SN~2006aj are similar 
to those of the non-GRB supernova SN~2002ap \cite{Foley2003}: small released energy and small ejected mass \cite{Mazzali2006}. These characteristics, along with the softness and the extremely 
long duration (2000 s, \cite{Barthelmy2006}) of GRB~060218,  suggest that this event is different from other 
GRB-SNe known so far. One critical parameter (see \cite{Mazzali2006} for a detailed modelling) could be 
the initial mass of the progenitor star: it may have been significantly
smaller for SN~2006aj ($\sim 20 M_{\odot}$) than  for other GRB-SNe ($\sim 40 M_{\odot}$). Moreover, the
collapse/explosion released less energy. A star of mass $\sim 20 M_{\odot}$ is expected to form a neutron
star rather than a black hole when its core collapses, possibly resulting in a highly magnetized neutron star (a magnetar) \cite{Mazzali2006}.

\subsection{The light curves}

\begin{figure}
  \includegraphics[height=.4\textheight]{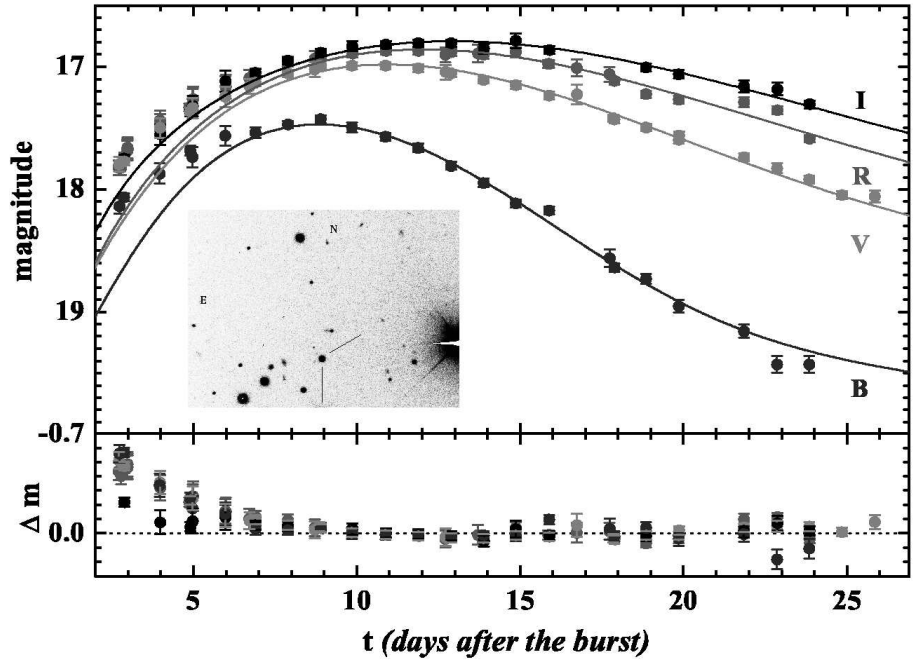}
  \caption{The BVRI light curves of SN~2006aj. The data have been corrected
for extinction and host flux contribution. The fit has been done using the light
curves of SN~1998bw as a template and not considering the data that were taken
before 8.8 days post burst. The residuals $\Delta m$ are the difference between the observed
values and the fit. The inset shows a VLT/FORS2 R-band image of SN~2006aj. Revised from \cite{Ferrero2006}.}
\label{figure2}
\end{figure}

Following \cite{Zeh2004,Zeh2005}, we modeled the light curves of SN~2006aj using SN~1998bw
as a template that was shifted to the corresponding redshift and scaled in luminosity (in the SN rest frame) by a factor $k$ and in time evolution by a factor $s$, while zero host extinction 
was assumed for SN~1998bw \cite{Patat2001}. In doing so, we find an
additional component visible in the early data that makes the light
curve systematically brighter than that of SN~1998bw (Fig. \ref{figure2}). This component
has also been noted by \cite{Mirabal2006b} and explained as the light due to a
shock break-out through a dense progenitor wind \cite{Campana2006}.

\smallskip

\begin{figure}
  \includegraphics[height=.5\textheight]{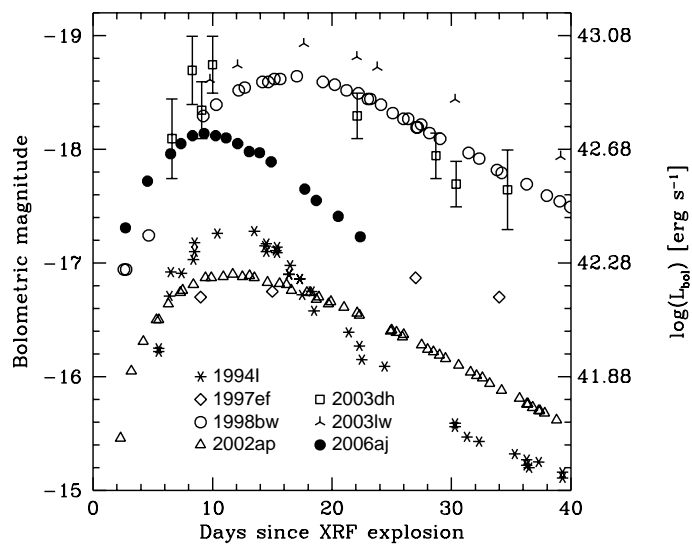}
  \caption{A comparison between the bolometric light curves of normal Type Ic SNe and GRB-SNe. 
  (For details see also \cite{Pian2006}.)}
\label{figure3}
\end{figure}

It is interesting to compare the properties of SN 2006aj with those of other Type Ic supernovae.
Fig. \ref{figure3} shows that the three well-observed, low-redshift GRB-SNe (SN~1998bw, SN~2003dh and SN~2003lw) are very similar with respect to their light curves. On the other hand, 
SN~2006aj at maximum light is fainter by about a factor of two, but it is still
a factor of 2-3 more luminous than the other broad-lined Type Ic supernovae not related to GRBs (see also
Fig. 6 in \cite{Ferrero2006}).

Normal Type Ic supernovae rise to a peak in approximately 10-12 days. Previously known GRB-SNe showed a longer rise time (14-15 days), but SN~2006aj
rose as fast as normal Type Ic SNe and also declined fast. 

SN~2006aj appears to be an intermediate object between GRB-SNe and other Type Ic SNe without a detected GRB.

\subsection{The host galaxy}

The host galaxy of GRB~060218/SN~2006aj was imaged pre-burst by the
Sloan Digital Sky Survey \cite{Cool2006}. These images were used to obtain its magnitudes \cite{Ferrero2006}, as the SN brightness was hiding the host component, and to measure its
diameter ($\sim 2$ kpc, for $z$=0.033).

\begin{figure}
  \includegraphics[height=.55\textheight]{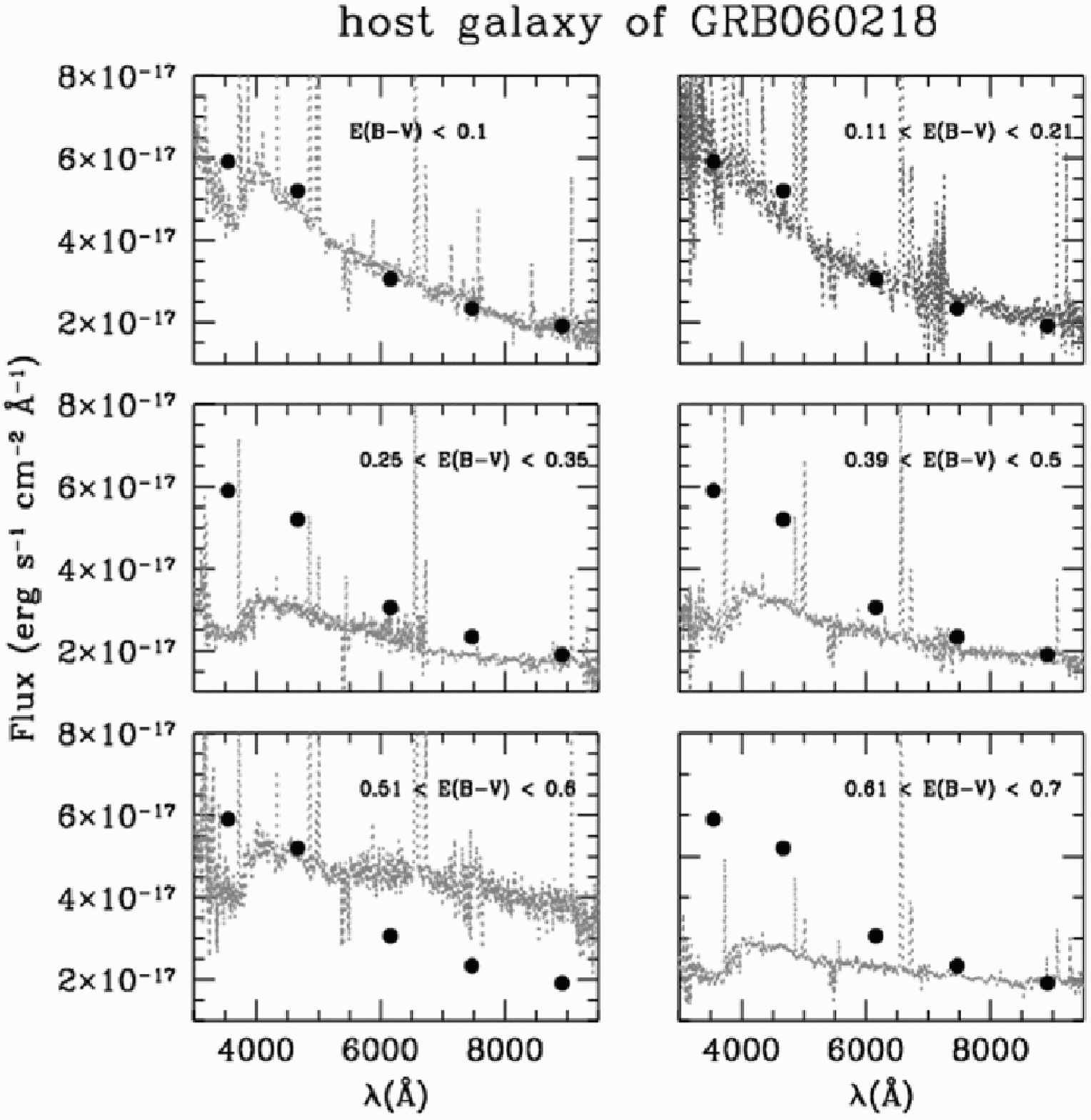}
  \caption{The broad-band SED of the host galaxy of SN~2006aj (thick points) is compatible with a   
  moderately absorbed galaxy (starburst templates from \cite{Kinney1996}). The data have been corrected
  for galactic extinction.}
\label{figure4}
\end{figure}

The VLT/UVES spectrum allowed a measurement of the total extinction toward SN~2006aj and it turned
out that the main contribution is from our Galaxy ($E(B-V)= 0.127 \pm 0.005$), while $E(B-V)_{host}$ is only $0.044 \pm 0.001$ \cite{Guenther2006}. The latter is supported by the SED of the galaxy (see 
Fig. \ref{figure4}).

In addition, spectroscopic studies led to an estimate of the $SFR \sim 0.07 \ M_{\odot}/year$ 
and of the metallicity $Z \sim 0.07 \ Z_{\odot}$ \cite{Wiersema2007}.
Taking into account slit losses, the $SFR$ might be understimated by a factor 3. This would imply
that the SN~2006aj host has a $SFR$ comparable to that of the Milky Way, while its stellar mass is 
about 1000 times smaller ($M_{host}=10^{7.2 \pm 0.3}$ $ M_{\odot}$) \cite{Savaglio}.

%%%%%%%%%%%%%%%%%%%%%%%%%%%%%%%%%%%%%%%%%%%%%%%%
%% BACKMATTER
%%%%%%%%%%%%%%%%%%%%%%%%%%%%%%%%%%%%%%%%%%%%%%%%

\begin{theacknowledgments}
  F.P. thanks Sylvio Klose for useful advice and comments and 
acknowledges support by DFG grant Kl 766/13-2 and by the German Academic Exchange
  Service (DAAD) under grant No. D/05/54048.
\end{theacknowledgments}

%%%%%%%%%%%%%%%%%%%%%%%%%%%%%%%%%%%%%%%%%%%%%%%%

\end{document}